\documentclass[conference]{IEEEtran}
\IEEEoverridecommandlockouts
\usepackage{cite}
\usepackage{amsmath,amssymb,amsfonts}
\usepackage{algorithmic}
\usepackage{graphicx}
\usepackage{textcomp}
\usepackage{xcolor}
\usepackage{hyperref}
\usepackage{multirow}

\title{Generating Privacy Stories From Software Documentation}

\author{\IEEEauthorblockN{Wilder Baldwin, Shashank Chintakuntla, Shreyah Parajuli, Ali Pourghasemi, Ryan Shanz, Sepideh Ghanavati}
\IEEEauthorblockA{\textit{School of Computing and Information Science} \\
\textit{University of Maine}\\
Orono, ME, United States \\
\{wilder.baldwin, shashank.chintakuntla,shreyash.parajuli, ali.pourghasemi, ryan.shanz, sepideh.ghanavati\}@maine.edu}
}

\begin{document}

\maketitle

\begin{abstract}
Research shows that analysts and developers consider privacy as a security concept or as an afterthought, which may lead to non-compliance and violation of users' privacy. Most current approaches, however, focus on extracting legal requirements from the regulations and evaluating the compliance of software and processes with them. In this paper, we develop a novel approach based on chain-of-thought prompting (CoT), in-context-learning (ICL), and Large Language Models (LLMs) to extract privacy behaviors from various software documents prior to and during software development, and then generate privacy requirements in the format of user stories. Our results show that most commonly used LLMs, such as GPT-4o and Llama 3, can identify privacy behaviors and generate privacy user stories with F1 scores exceeding 0.8. We also show that the performance of these models could be improved through parameter-tuning. Our findings provide insight into using and optimizing LLMs for generating privacy requirements given software documents created prior to or throughout the software development lifecycle.     
\end{abstract}
\vspace{-2.0mm}

\section{Introduction}
\label{sec:intro}


Understanding the privacy behaviors of software applications and eliciting privacy requirements during the early phases of the software development lifecycle (SDLC) are essential for developing privacy-preserving and regulatory-compliant software \cite{gurses2016privacy, ceross2024rethinking}. Past research, however, shows that software analysts and developers often consider privacy as a subset of security requirements or as an afterthought \cite{hadar2017privacy, prybylo2024evaluating}, and they often lack the tools needed to understand and identify privacy behaviors of the applications they develop \cite{jain2022pact, jain2023towards}. 

Most common approaches for identifying and eliciting privacy requirements include conducting privacy impact assessments \cite{deng2011privacy, canedo2023privacy}, or employing goal-oriented methodologies to map privacy requirements to system processes \cite{ghanavati2014legal, canedo2023privacy, kalloniatis2008addressing}. Other works aim to extract privacy-related information from user stories or use case models \cite{herwanto2022privacystory, casillo2022detecting, herwanto2024leveraging, bartolini2019gdpr, mai2018modeling, ronanki2023investigating, krishna2024using} by leveraging Natural Language Processing (NLP) techniques and then using predefined templates to generate privacy requirements. However, these approaches mostly focus on the specific forms of software documentation (i.e., user stories or use cases), or they rely on developers to understand how personal information is handled by their applications. In addition, they require a large amount of labeled data for such identification and generation. 


In recent years, prior work has applied Large Language Models (LLMs) for tasks such as textual documents classification and generation ~\cite{brown2020language, wei2022chain, xiao2025foundations, tan2024large},  requirements verification~\cite{santos2024requirements}, privacy requirements analysis~\cite{rodriguez2024large, ronanki2023investigating}, and GDPR-related requirements generation (such as Record of Processing Activities) \cite{pragyan2024toward}. Inspired by these prior work, in this paper, we develop a novel framework that leverages prompt engineering and post-training techniques to help detect and extract software applications' privacy behaviors (i.e., actions, data types, and purposes) from various unstructured software artifacts such as software requirements and code specification documents, and architectural design documents and generate privacy user stories (in a short form \textit{privacy stories}). Our dataset represents real-world software documents created throughout the SDLC. While it is important to identify requirements early in development, privacy requirements may dynamically evolve with the software, especially in an agile environment. We hypothesize that LLMs would be an ideal approach to help create and update privacy requirements and, thus, test our framework on software documents created during various stages of this process. We do not use privacy policies or privacy labels since past research has shown that there are inconsistencies between privacy policies, privacy labels, and the software applications \cite{yu2018ppchecker, okoyomon2019ridiculousness, zimmeck2019maps, li2022understanding, Khandelwal2024, jain2023atlas} or the developers over-report or under-report privacy labels \cite{Khandelwal2024}. 

\begin{figure}
    \centering
    \includegraphics[width=\columnwidth]{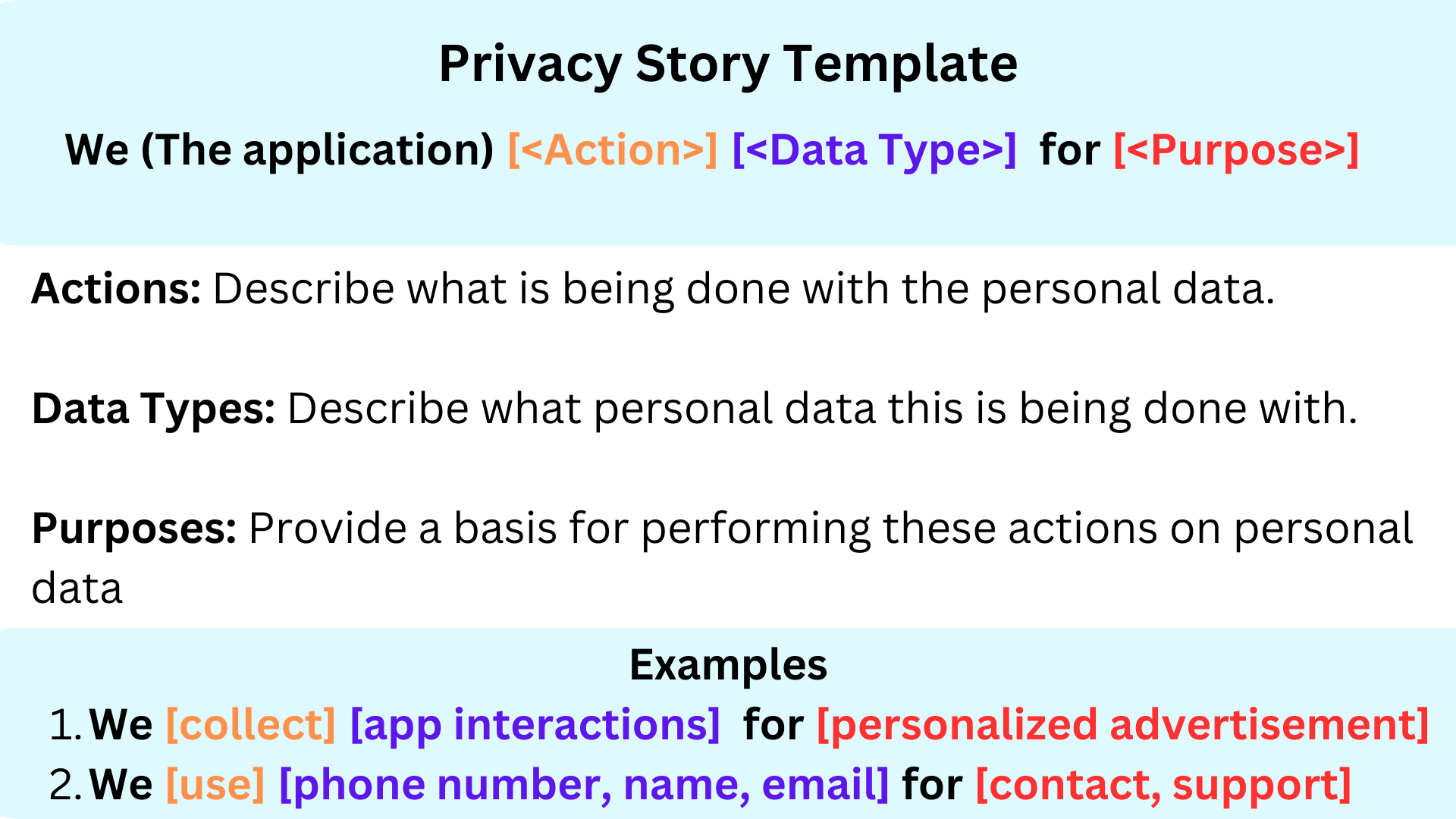} 
    \caption{An Example of a Privacy Story}
    \label{fig:privacy_story}
\end{figure}

We collect files from 25 sets of software documents that describe apps' interaction with personal information from 20 open-source repositories. We extend an already existing privacy action taxonomy, called PAcT \cite{jain2022pact}, with 3 actions, 50 data types, and 26 purpose labels, and define a simplified template for privacy requirements in terms of privacy stories (see Fig. \ref{fig:privacy_story}). We then annotate these documents to identify their privacy behaviors and create privacy stories. Our dataset includes 170 privacy behaviors and 93 privacy stories that are manually created and evaluated. Lastly, we develop a novel framework that combines prompt engineering approaches with parameter-tuning techniques, such as thought preference optimization (TPO) \cite{wu2024thinking}, to automate the process. We evaluate various popular commercial and open-source LLMs' performance for the tasks, quantitatively and qualitatively, with two independent annotators. Our research questions are:






\textbf{- RQ1:} How accurately can LLMs extract privacy behaviors and generate privacy stories from software documents?

In our quantitative analysis, the best-performing models, Llama-3.3-70b and GPT-4o, exceed 0.8 F1 scores for extracting actions, data types, and purposes as privacy behaviors. Similarly, these models generate privacy stories with more than 80\% precision. According to our annotators, only 16\% of these accurate stories could be improved.

\textbf{- RQ2:} What are the LLMs' challenges to accurately generate privacy stories given software documents? 

In our quantitative and qualitative analysis, we found that using predefined labels, particularly for data types, contributed to lower F1 scores, as models often generated new labels or incorrectly applied existing ones. This evaluation revealed that GPT-4o is more conservative in extracting privacy behaviors, with a higher precision and a lower recall than the open-source models we tested. A human evaluation of LLM-generated privacy stories revealed a tendency to hallucinate new labels.




\section{Background}
\label{sec:background}

We outline key terms and background used in this paper. 

\subsection{Privacy Action Taxonomy (PAcT)}
\label{sec:pact}

Privacy Action Taxonomy (PAcT) \cite{jain2022pact} classifies privacy behaviors in source code in terms of practices/actions (e.g., collection), data types (e.g., location data), and purposes (e.g., functionality).  We initially extend PAcT using labels from the Google Play Store’s data safety section \cite{google2023data}. We, then, follow an open coding procedure to expand this initial taxonomy with all privacy behaviors extracted from a dataset of 25 software documents. Fig. \ref{fig:pact_outline} shows our extended PAcT taxonomy. 

\subsection{Privacy (User) Stories}
\label{sec:privacy-user-story}

User stories are a popular form of representing software requirements  \cite{cohn2004user, herwanto2024leveraging}. We extend this concept with a specialized form that focuses on privacy requirements, called privacy (user) stories or privacy stories, in short (see Fig. \ref{fig:privacy_story}). 
We use a simplified privacy story template that takes in actions, data types, and purposes regarding personal data in the format: ``We (i.e., The Application) (action) (data type) for (purpose)''. 

\subsection{Prompt Engineering and Parameter Tuning}
\label{sec:prompting}

In-context learning (ICL) refers to LLMs' abilities to recognize tasks, often in the form of examples, patterns, and analogies at inference time \cite{brown2020language}. Chain-of-thought (CoT) prompting \cite{wei2022chain} has inspired new approaches to use LLMs for reasoning problems~\cite{xia2024beyond, plaat2024reasoning}. Similar to CoT, other frameworks are proposed to improve and evaluate LLMs' ability to judge and verify their outputs \cite{jacovi2024chain, dhuliawala2023chain}, reduce hallucinations \cite{du2024haloscope}, and retrieve task-relevant context from datasets \cite{wang2024astute}. These approaches could also be paired with model tuning approaches such as instruction tuning \cite{ziegler2020fine}, 
to enable LLMs to complete complex tasks that require multi-step reasoning \cite{zhang2023igniting}. 

Supervised Fine-Tuning (SFT) is a parameter-tuning technique used to adapt LLMs to specific tasks by training them on a labeled dataset \cite{ziegler2020fine}. In SFT, the model is presented with input-output pairs, where the input is a prompt or document snippet, and the output is the expected response (e.g., software documents and our annotated privacy behaviors, as well as stories are the input-output pair). 

Direct Preference Optimization (DPO) is an LLM training and policy alignment technique that updates the model's parameters by optimizing a binary cross-entropy objective, which learns a reward function from preferred and dispreferred responses to a given prompt \cite{rafailov2023dpo}.


In this paper, we leverage several prompt engineering and parameter-tuning techniques to evaluate the performance of LLMs on privacy behavior extraction and privacy story generation tasks. Two key methods we employ for parameter-tuning are SFT and DPO, which are critical for optimizing language models for our specific use case.

 
\section{Related Work}
\label{sec:related}

In recent years, various frameworks and approaches have been introduced to integrate and evaluate LLMs within the SDLC, particularly for tasks such as requirements generation and refinement \cite{lubos2024leveraging, ijetajournal2024, lin2024when, arora2024advancing}. Zhang et al. \cite{zhang2024llm} and Krishna et al. \cite{krishna2024using} propose agentic frameworks that leverage LLMs to enhance user stories, and evaluate their effectiveness through structured tasks and contextual information extraction. They also conducted a study with developers to analyze these enhancements, and they found a trade-off between ambiguity and over-complexity in requirements when using more powerful models. Ronanki et al. \cite{ronanki2023investigating} investigate ChatGPT’s capabilities, strengths, and shortcomings in generating software requirements. They found that its responses scored higher across various quality attributes compared to those of the developers. 

Extracting requirements from software documents has been an application of NLP tools and recently LLMs. Mu et al. \cite{mu2009extracting} propose an approach to extract functional requirements from free-text documents. They introduce an extended functional requirements framework comprising ten semantic cases to capture the linguistic characteristics of Software Requirements Specifications (SRSs). Slankas et al. \cite{slankas2013automated} develop an approach to automate NFR extraction by parsing the document and then classifying the result into 14 types of NFRs. They observe that support vector machine and k-nearest neighbor approaches achieve the highest accuracy. 
Ferrari et al. \cite{ferrari2017pure} fine-tune a BERT model on a manually annotated dataset derived from the PURE corpus \cite{ivanov2022extracting}, to identify technical requirements within natural language texts. They achieve a precision of 92\% and a recall of 80\% on the PURE dataset. Lubos et al. \cite{lubos2024leveraging} show that LLMs are effective at spotting quality issues and improving requirements in a collaborative environment with developers and analysts. These findings are complemented by \cite{krishna2024using}, which found LLMs could aid entry-level developers through the creation of software requirement specification documents. 

LLMs have been used both as annotators \cite{tan2024large, wang2024human,li2024comparative} and for analyzing privacy policies, software documents, and code \cite{rodriguez2024large, arora2024advancing, morales2024large}. Rodriguez et al. \cite{rodriguez2024large} use LLMs to extract data types from privacy policies through a variety of ICL-focused prompt engineering and parameter tuning approaches.  Their approach achieves comparable F1 scores to other state-of-the-art approaches, such as support vector machine \cite{wilson2016creation} in identifying personal data types.
Morales et al. \cite{morales2024large} compare various LLMs to extract data practices from privacy policies and detect misalignments between app behaviors gathered from taint analysis and policies. Tang et al. \cite{tang2023policygpt} introduce the PolicyGPT, a framework for leveraging LLMs for automated privacy policy analysis. They test PolicyGPT on OPP-115 (115 website policies) and PPGDPR (304 mobile app policies). PolicyGPT achieves 97\% and 87\% accuracy under zero-shot conditions, outperforming traditional NLP models. Woodring et al. \cite{wei2022chain} develop Privacify, a user-centric web application, designed to enhance the accessibility and understanding of privacy policies. Their work uses LLMs to perform text segmentation and summarization. It also simplifies complex legal jargon into clear, user-friendly language with high reliability and accuracy, achieving an accuracy of 90.5\%. 

In this paper, we extend the evaluation of LLMs by examining their effectiveness in analyzing software documentation for privacy behaviors and generating privacy-related requirements.


\begin{figure}
    \centering
    \includegraphics[width=\columnwidth]{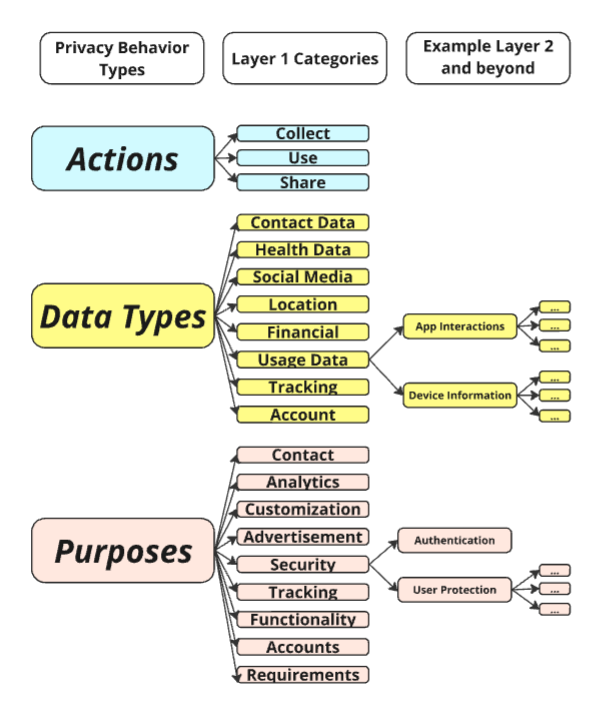} 
    \caption{An Excerpt of Our Extended Privacy Action Taxonomy}
    \label{fig:pact_outline}
\end{figure}

\section{Privacy Stories Generation Framework}
\label{sec:methodology}

\begin{figure}
    \centering
    \includegraphics[width=\columnwidth]{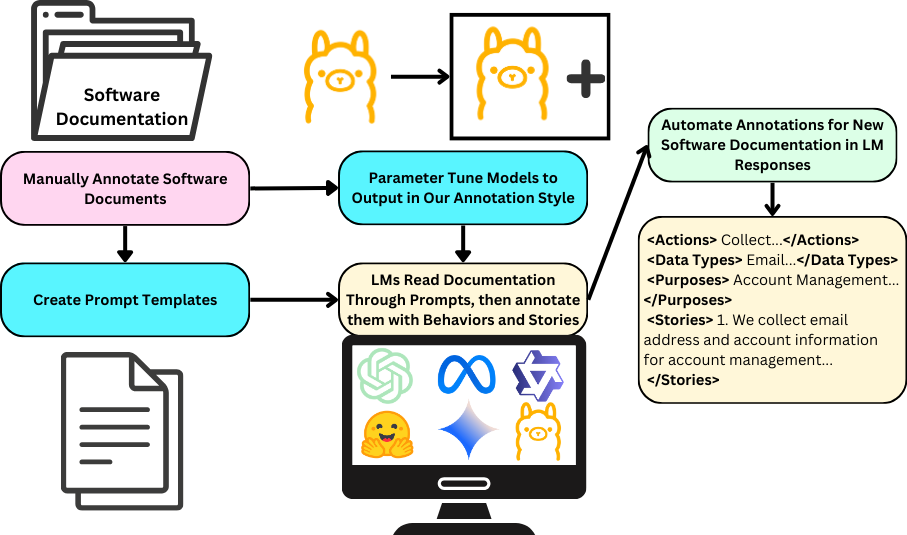} 
    \caption{Overview of Our Framework for Generating Privacy Stories}
    \label{fig:process_flow}
\end{figure}

We develop a framework to help extract privacy behaviors (e.g., actions, data types, or purposes) from software documents and then generate privacy requirements in the form of privacy (user) stories. 
The framework consists of three main components (as shown in Fig. \ref{fig:process_flow}): (1) ground-truth dataset creation (i.e., Step 1); (2) prompt template design (i.e., Step 2); and (3) LLM-based privacy story generation (i.e., Step 3). In \textit{Step 1}, we create a ground-truth dataset of the publicly available software documents and manually annotate them with their privacy behaviors and write privacy stories (see Section \ref{sec:background} for their definition). In \textit{Step 2}, we design a prompt template to extract privacy behaviors and generate privacy stories. To achieve this, we evaluate various prompt engineering approaches and examine multiple prompt templates on a new set of software documents to ensure the generated privacy stories align with our dataset in Step 1.
In \textit{Step 3}, we investigate several human feedback learning methods for training various LLMs to identify privacy behaviors and generate privacy stories; hence, improving the accuracy. 


\subsection{Ground-Truth Dataset Creation}
In Step 1, we create a ground-truth dataset of software documents annotated with privacy behaviors and privacy stories to use for LLMs training. 

Given the lack of available open-source software documents' datasets, we gather documents from several sources to create our ground-truth. First, we crawled GitHub open-source repositories to extract folders with naming conventions, such as  ``documentation'' or ``docs''. Then, the first three authors  reviewed the results to identify documents that describe how an application interacts with personal data. The data returned was sparse, and difficult to find examples of real-world applications handling personal data. To speed up our search, we examined the F-Droid marketplace \cite{fdroid} to extract documents from open-source Android applications. Additionally, we inferred LLMs to provide links from their pre-training data, and from web search tools to identify open-source software documents. In total, we collected 25 sets of software documents from 20 unique applications to use for the annotation. Our set consists of various forms; some describe system architectures and data models in UML format, others include developer, user, and API guides for applications, or general descriptions of the application's requirements and code. A variety of application documents are included in our dataset, e.g., health-tech, mobile, and financial applications.  



After collecting the data, the same authors reviewed each document individually to identify the actions, data types, and purposes. The result is then used in the privacy story template to write privacy stories (see Section \ref{sec:privacy-user-story}). 
To label behaviors and create the dataset of privacy stories, we extended the Privacy Action Taxonomy (PAcT)~\cite{jain2022pact} (see Section \ref{sec:pact}). The Google Play Store requires developers to disclose collection, use, and sharing of data types and their purposes through standardized labels \cite{google2023data}. We use these labels as the initial set for the annotation task. However, throughout the annotation process, if the annotators could not map the text to any of the labels, they added a new label and continued to do so until they reached saturation \cite{glaser2002concept}. In addition, they used the privacy story template to write privacy stories based on the labels. After independently annotating all the documents and writing the relevant privacy stories, the second and third authors discussed their results together to consolidate the labels and privacy stories and to reach an agreement. Since the number of annotations, the labels, and the privacy stories were large, we did not calculate the Kappa agreement metrics. Next, the first author independently reviewed and compared the labels and the privacy stories and finalized the taxonomy (shown in Fig. \ref{fig:pact_outline}). At the end, the extended PAcT includes 3 actions, 50 data types, and 26 purpose labels, and our ground-truth dataset consists of 171 privacy behaviors (i.e., 50 actions, 60 data types, and 61  purposes) and 93 privacy stories. 





\subsection{Prompt Template Design}

In Step 2, we design the optimal prompt template that suits the tasks of identifying privacy behaviors and generating privacy stories (by filling out the template) from software documents. We leverage the Chain-of-Thought (CoT) prompting approach \cite{wei2022chain} to break down the tasks into smaller steps and In-Context-Learning (ICL) \cite{brown2020language} to ensure that the model learns from prior examples and follows a structured output. 

\begin{figure}
    \centering
    \includegraphics[width=\columnwidth]{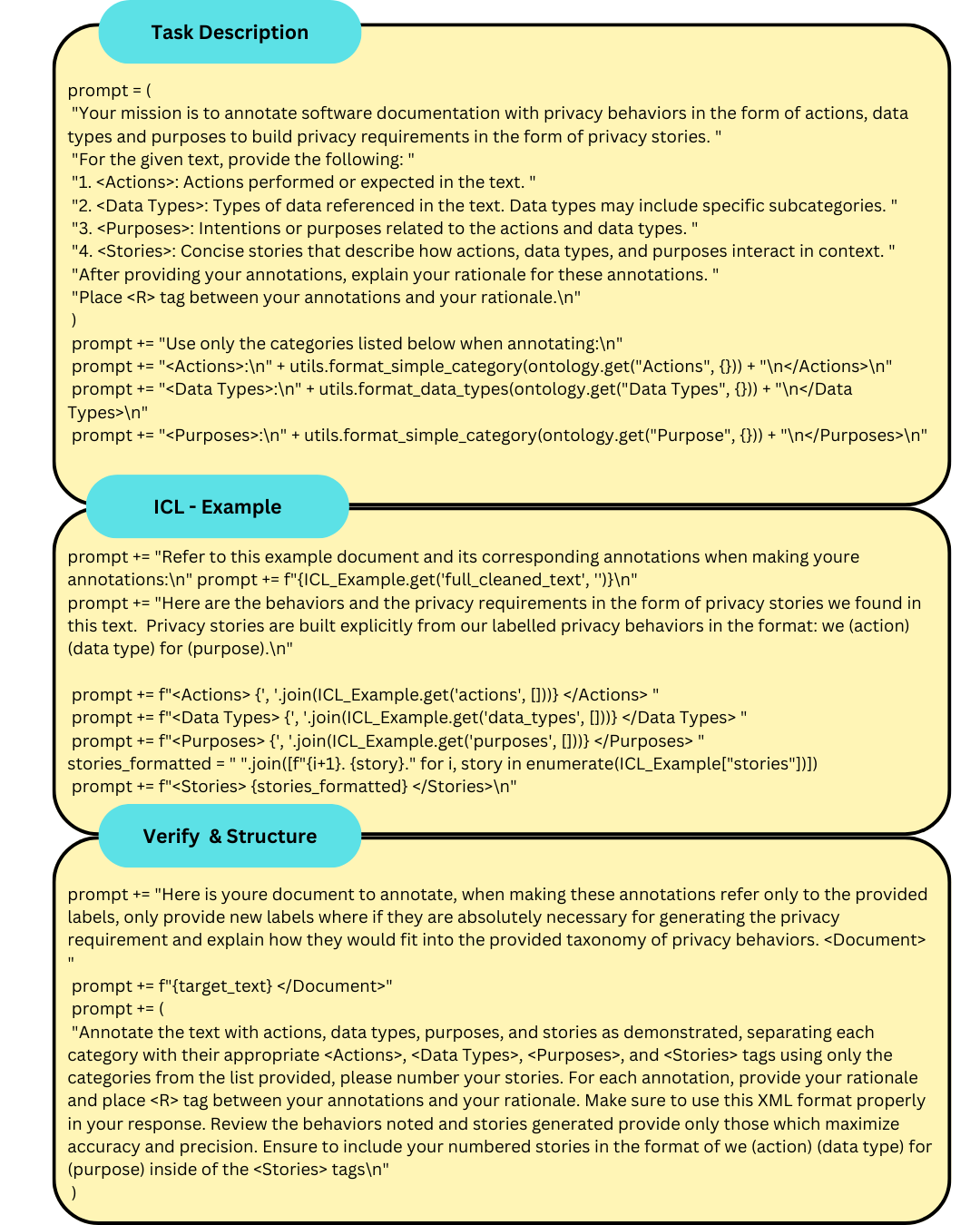} 
    \caption{LLM Annotation Prompt Template}
    \label{fig:prompt_template}
\end{figure}

Our base prompt template includes a task description, with the extended PAcT labels, and the document to annotate. 
We extend this base prompt with three additional features based on the prior research. 
First, we provide an ICL example, which is a document from our ground-truth dataset with its privacy behaviors and stories. To find the best example document, we use sentence transformers \cite{reimers2019sentence} to generate embeddings for all 25 documents. We then compute cosine similarity over these embeddings to select the document that is most semantically aligned with the input document. Second, we provide instructions on how to format the responses (i.e., output). To accomplish this, we follow insights from prior research \cite{nguyen2025highlighted, wu2024thinking} that show that using XML tags within prompts may improve the structure for formatting model inputs and outputs.  
Thus, our prompt templates instruct the model to format actions, data types, purposes, stories, and the rationale with XML tags that we can use to parse and evaluate in Step 3 (i.e., Section \ref{sec:LM-story}). Third, in the final step of our prompt template design, we restate the task description and instruct the model to provide a rationale and verify its outputs. Providing rationale and verification are common strategies to increase model accuracy \cite{zhao2023survey} and to provide more comprehensive training data to help refine thinking patterns for other LLMs \cite{ wu2024thinking}. Fig. \ref{fig:prompt_template} shows our final prompt template. 

Lastly, we decided to keep the default temperature value (0.7) for all the models to ensure creating more diverse training data. Previous research shows that reducing the default temperature value to 0.0 could improve privacy analysis~\cite{rodriguez2024large}, and requirements elicitation \cite{santos2024requirements}. In our work, however, all LLM outputs are reviewed by two authors.  We hypothesize that fine-tuning LLMs for this task can reduce variability; hence, we keep the temperature value as-is. 


\subsection{LLM-based Privacy Story Generation}
\label{sec:LM-story}

To make the most of our small dataset with only 25 software documents, we follow guidelines regarding optimizing LLMs for annotating \cite{tan2024large} and general LLM finetuning approaches \cite{xiao2025foundations, zhao2023survey}. Following these guidelines, in Step 3, we optimize LLM responses through supervised fine-tuning (SFT), and then continue refinement through policy alignment \cite{ziegler2020fine, rafailov2023dpo} (see Section \ref{sec:prompting} for their description). 


In the first SFT stage, we enhance our ground-truth dataset. By combining our base prompt template as input with our annotations as output, this data can be used for performing SFT. The goal of the SFT step is to align model parameters toward outputting similar labels to our manually defined labels for a given software document. 

We further investigate training LLMs by drawing insights from Thought Preference Optimization (TPO) \cite{wu2024thinking}. This framework utilizes DPO \cite{rafailov2023dpo} with a preference choice dataset of LLM responses that includes reasoning, to align models' thinking patterns towards correct outputs. As suggested for TPO \cite{wu2024thinking}, our prompt templates instruct LLMs to separate their outputs (i.e., privacy behaviors and stories) and reasoning with an \texttt{<R>} tag. 

The fourth and fifth authors, who were not involved in the labeling process, individually reviewed two outputs for each document from the top-performing model (i.e., Llama-3.3-70b) to create a preference choice dataset. We utilize this dataset to train our LLMs' thinking patterns using DPO. We report the results of this training step in Section \ref{sec:parameter-tuning}.



To address the high computational cost of parameter-tuning LLMs, we adopt the following strategies that enable us to complete training in a free Google Colab environment (one x T4 GPU). We employ DPO with Unsloth FastLanguageModel for efficient fine-tuning \cite{unsloth2024unsloth}. Our configuration uses a batch size of 2, gradient accumulation of 3, and a learning rate of 5e-6, training for 3 epochs. We apply LoRA with a rank of 64, targeting transformer modules to adapt the Llama-3.1-8b and Llama-3.2-3b models. We choose a beta value of 0.1, controlling preference alignment in DPO training using the \texttt{trl} library \cite{huggingface2024transformers}. 



\section{Results}
\label{sec:result}

We report the results of privacy behavior identification, privacy story generation, as well as prompt engineering and model-tuning approaches. We conduct the analysis with both open-source and proprietary LLMs that are widely used in industry and research, and achieve the state-of-the-art performance on popular benchmarks at their time of release. The LLMs we use are: GPT-4o \cite{openai2023gpt}, as well as open source models from the Llama-3  \cite{dubey2024llama}, and Qwen 2.5 families \cite{yang2025qwen2.5}. 

\begin{table*}[htbp]
\caption{F1 Scores For Base Model Annotations (All Files)}
\centering
\resizebox{\textwidth}{!}{
\begin{tabular}{|l|c|c|c|c|c|c|c|c|c|c|c|c|}
\hline
\textbf{Model} & \textbf{Overall F1} & \textbf{Overall Precision} & \textbf{Overall Recall} & \textbf{A F1} & \textbf{A Precision} & \textbf{A Recall} & \textbf{DT F1} & \textbf{DT Precision} & \textbf{DT Recall} & \textbf{P F1} & \textbf{P Precision} & \textbf{P Recall} \\  \hline
GPT4o & 0.722 & 0.744 & 0.715 & \textbf{0.930} & 0.951 & 0.929 & 0.505 & 0.485 & 0.564 & 0.7368 & 0.703 & 0.795 \\ \hline
Qwen-2.5-32b & 0.727 & 0.701 & 0.771 & 0.883 & 0.844 & 0.942 & 0.581 & 0.516 & 0.703 & 0.684 & 0.742 & 0.667 \\ \hline
Deepseek-R1-Distill-Qwen-32b & 0.559 & 0.564 & 0.566 & 0.689 & 0.694 & 0.684 & 0.380 & 0.341 & 0.446 & 0.597 & 0.662 & 0.557 \\ \hline
Llama3.3-70b & \textbf{0.766} & 0.737 & 0.810 & 0.864 & 0.813 & 0.938 & \textbf{0.641} & 0.597 & 0.732 & \textbf{0.762} & 0.801 & 0.761 \\ \hline
Llama3.1-8b & 0.601 & 0.564 & 0.671 & 0.676 & 0.648 & 0.724 & 0.496 & 0.431 & 0.661 & 0.607 & 0.613 & 0.628 \\ \hline
LLama3.2-3b & 0.372 & 0.342 & 0.422 & 0.324 & 0.311 & 0.347 & 0.347 & 0.294 & 0.499 & 0.409 & 0.421 & 0.420 \\ \hline
\end{tabular}
}
\label{tab:f1_overall-behaviors}
\end{table*}

\begin{table*}[h!]
\caption{F1 Scores For Base Model Annotations by File Type}
\centering
\resizebox{\textwidth}{!}{
\begin{tabular}{|l|c|c|c|c|c|}
\hline
\textbf{Model} & \textbf{Overall F1}  & \textbf{Software \& Code Spec F1 (8)} & \textbf{User \& Developer Guides F1 (9)} & \textbf{Architecture \& DB Design Diagrams F1 (6)} & \textbf{README F1 (2)}  \\ \hline
GPT4o & 0.722 & 0.725 & \textbf{0.752} & 0.695 & 0.627 \\ \hline
Qwen-2.5-32b & 0.727 & 0.804 & 0.711 & \textbf{0.762} & 0.621  \\ \hline
Deepseek-R1-Distill-Qwen-32b & 0.559 & 0.566 & 0.536 & 0.608 & 0.571 \\ \hline
Llama3.3-70b & \textbf{0.766} & \textbf{0.832} & 0.738 & 0.711 & \textbf{0.778}  \\ \hline
Llama3.1-8b & 0.601 & 0.639 & 0.549 & 0.625 & 0.631 \\ \hline
LLama3.2-3b & 0.372 & 0.277 & 0.451 & 0.514 & 0.0 \\ \hline
\end{tabular}
}
\label{tab:f1_overall-filetypes}
\end{table*}

\begin{table}[h]
\caption{F1 Scores For Base Prompt Model Annotations}
\centering
\resizebox{\columnwidth}{!}{
\begin{tabular}{|l|c|c|c|}
\hline
\textbf{Model} & \textbf{Overall F1} & \textbf{Overall Precision} & \textbf{Overall Recall} \\ \hline
GPT4o & 0.288 & 0.212 & 0.335 \\ \hline
Qwen-2.5-32b & \textbf{0.428} & 0.375 & 0.516 \\ \hline
Deepseek-R1-Distill-Qwen-32b & 0.161 & 0.161 & 0.180 \\ \hline
Llama3.3-70b & 0.239 & 0.222 & 0.259 \\ \hline
Llama3.1-8b & 0.034 & 0.029 & 0.052 \\ \hline
LLama3.2-3b & 0.074 & 0.050 & 0.147 \\ \hline
\end{tabular}
}
\label{tab:f1_base_prompts} 
\end{table}

\subsection{Analysis of Privacy Behaviors' Extraction Task}
\label{sec:behavior-model}

We analyze the accuracy of LLMs for the task of extracting privacy behaviors from software documents by comparing their results with our annotated dataset, which serves as the gold labels. We use the F1 score based on the definition proposed by \cite{kiritchenko2006learning}, which is suitable for hierarchical data, similar to ours. 
Based on their method and to account for the hierarchy in our labels, if an extracted behavior is found on the tree of a true/gold label, we divide the score by its distance on the tree from that true label. Note that the tree begins with the first layer of the taxonomy. For example, the label \textit{``Usage Data''} receives a 0.5 score if the true label is \textit{``App Interactions''}, since it is the parent of the \textit{``Usage Data''} in our taxonomy. When parsing LLM responses, we exclusively gather labels that match the labels in our taxonomy. 

Table \ref{tab:f1_overall-behaviors} shows the precision, recall, and F1 scores for Action (A), Data Types (DT), and Purpose (P) for all LLM models mentioned above. Llama-3.3-70b scores the highest overall F1 (0.766) outperforming Qwen-2.5-32b (0.727), and GPT-4o (0.722). Except for GPT-4o, recall is higher than precision, which may indicate that this larger model (i.e., GPT-4o) is more conservative in its identification of privacy behaviors than the open-source models. Except for Llama3.2-3b, the models have significantly higher F1 scores for actions than data types and purposes, with the highest F1 score of 0.930. This is probably because the data types and purposes in our taxonomy consist of more categories and more layers (i.e., depth) in comparison to action privacy behaviors, which makes them harder to predict. Llama3.3-70b outperforms the other models in detecting data types and purposes with F1 scores of 0.641 and 0.762, respectively. Table \ref{tab:f1_base_prompts} shows the overall performance (i.e., F1 score, precision, and recall) of our base prompt outlined in Section \ref{sec:methodology} for all the models. Comparing the overall F1 scores of the base prompt with the final prompt (i.e., Table \ref{tab:f1_overall-behaviors} with Table \ref{tab:f1_base_prompts}) shows that the F1 scores are improved significantly using our final prompt (i.e., Fig. \ref{fig:prompt_template}) with the increase of between 0.299 - 0.722 across various models.


We analyze the accuracy of different models in identifying privacy behaviors from each type of document. 
Our dataset includes various types of unstructured documents that are commonplace in software development today~\cite{alzahrani2024software}. Table \ref{tab:f1_overall-filetypes} shows this distribution across different types of software documents for the LLMs we evaluated. The F1 scores represent the average score across all files of each type from the hierarchical F1 calculation laid out in Section \ref{sec:behavior-model}. Llama3.3-70b outperforms the other models across all documents (F1 = 0.766). Overall, the software requirements \& code specification documents perform better than the other types of documents, with the highest performance of F1 = 0.832 for Llama3.3-70b. For user developer guides and architecture, and DB design diagrams, GPT4.0 and Qwen-2.5-32b outperform Llama3.3-70b (F1 = 0.752 and F1 = 0.762, respectively) while LLama3.3-70b outperforms the rest of the models for README files. Similar to the previous analysis, the smaller Llama model (Llama3.2-3b) performs the worst across all documents. The variety in the results and the low F1 scores indicate the complexity of these documents. 

\subsection{Analysis of Privacy Story Generation Task} 
\label{sec:story-qual}

To evaluate the privacy stories generated by our approach, we use human annotators (i.e., authors 4-5) to manually review privacy stories generated by LLMs for their correctness and precision. The annotators were provided documentation files in text format, with their original manual annotations and the LLM-generated privacy stories. Each annotator answers the following questions for each generated privacy story: \textit{Q1. Is this accurate? (Yes/No)}, and \textit{Q2. Is this story missing any behaviors? (Yes/No)}. For the stories generated for one single document, the annotators also answer \textit{Q3. What stories are missing? (Text entry)}. 

The annotators individually reviewed the privacy stories generated by Llama3.3-70b, our highest performing model (see Section~\ref{sec:behavior-model}). The results for Q1 are shown in Table \ref{tab:annotators_q1_accurate_story}. Out of the 120 privacy stories generated across all 25 files by our model, 88 (73.3\%) were marked as accurate by at least one of the two annotators. 83 out of the 120 were marked as inaccurate by at least one of our annotators, where 32 (26.7\%) of them included hallucinated behaviors that were not found in PAcT. Of these 88 marked as accurate, 38 of them miss privacy behaviors according to at least one annotator (i.e., Q2). For Q3, annotators identified that 36 out of 93 privacy stories that existed in our original dataset were not generated (i.e., missing) by LLMs, which is $\sim$61\% recall.  


\begin{table}[h]
\centering
\caption{Annotators' Responses to Q1 - for All Generated Stories}
\resizebox{\columnwidth}{!}{%
\begin{tabular}{|l|rr|rr|rr|rr|rr|}
\hline
\multirow{2}{*}{} & \multicolumn{2}{c|}{\textbf{Overall}} & \multicolumn{2}{c|}{\textbf{Code Spec (8)}} & \multicolumn{2}{c|}{\textbf{User Guides (9)}} & \multicolumn{2}{c|}{\textbf{Arch/DB (6)}} & \multicolumn{2}{c|}{\textbf{README (2)}} \\
\cline{2-11}
 & \textbf{Yes} & \textbf{No} & \textbf{Yes} & \textbf{No} & \textbf{Yes} & \textbf{No} & \textbf{Yes} & \textbf{No} & \textbf{Yes} & \textbf{No} \\
\hline
\textbf{Both Annotators} & 41 & 27 & 20 & 9 & 19 & 12 & 2 & 4 & 0 & 2 \\
\hline
\textbf{At Least One Annotator} & 88 & 83 & 32 & 29 & 43 & 36 & 12 & 15 & 1 & 3 \\
\hline
\end{tabular}}
\label{tab:annotators_q1_accurate_story}
\end{table}

\begin{table*}[htbp]
\caption{F1 Score over test data including parameter tuned models}
\centering
\resizebox{\textwidth}{!}{
\begin{tabular}{|l|c|c|c|c|c|c|c|c|c|c|c|c|}
\hline
\textbf{Model} & \textbf{Overall F1} & \textbf{Overall Precision} & \textbf{Overall Recall} & \textbf{A F1} & \textbf{A Precision} & \textbf{A Recall} & \textbf{DT F1} & \textbf{DT Precision} & \textbf{DT Recall} & \textbf{P F1} & \textbf{P Precision} & \textbf{P Recall} \\  \hline
GPT4o & 0.772 & 0.780 & 0.780 & 0.930 & 0.933 & 0.944 & 0.704 & 0.681 & 0.768 & 0.651 & 0.725 & 0.628 \\ \hline
Qwen-2.5-32b & 0.806 & 0.766 & 0.865 & 0.925 & 0.889 & 0.978 & 0.719 & 0.624 & 0.864 & 0.754 & 0.785 & 0.754 \\ \hline
Deepseek-R1-Distill-Qwen-32b & 0.620 & 0.651 & 0.603 & 0.755 & 0.800 & 0.722 & 0.479 & 0.454 & 0.517 & 0.619 & 0.697 & 0.569 \\ \hline
Llama3.3-70b & 0.825 & 0.805 & 0.864 & 0.950 & 0.911 & 1.0 & 0.754 & 0.719 & 0.858 & 0.825 & 0.805 & 0.864 \\ \hline
Llama3.1-8b & 0.701 & 0.672 & 0.753 & 0.838 & 0.789 & 0.900 & 0.586 & 0.537 & 0.699 & 0.657 & 0.692 & 0.661 \\ \hline
LLama3.2-3b & 0.297 & 0.260 & 0.350 & 0.178 & 0.167 & 0.168 & 0.421 & 0.342 & 0.594 & 0.273 & 0.260 & 0.290 \\ \hline
Llama-3.1-8b + SFT & 0.766 & 0.755 & 0.801 & 0.938 & 0.889 & 1.000 & 0.677 & 0.656 & 0.754 & 0.655 & 0.720 & 0.650 \\ \hline
Llama-3.1-8b + SFT + TPO & 0.753 & 0.785 & 0.741 & 0.910 & 0.911 & 0.933 & 0.709 & 0.747 & 0.696 & 0.619 & 0.698 & 0.594 \\ \hline
Llama-3.2-3b + SFT & 0.751 & 0.755 & 0.752 & 0.887 & 0.900 & 0.878 & 0.725 & 0.719 & 0.735 & 0.634 & 0.646 & 0.643 \\ \hline
Llama-3.2-3b + SFT + TPO & 0.467 & 0.482 & 0.457 & 0.735 & 0.778 & 0.711 & 0.386 & 0.373 & 0.422 & 0.262 & 0.296 & 0.237 \\ \hline
\end{tabular}
}
\label{tab:f1_overall-models}
\end{table*}

\subsection{Parameter-Tuning Analysis}
\label{sec:parameter-tuning}


Table \ref{tab:f1_overall-models} displays the performance of base LLMs and a fine-tuned Llama 3.1-8b, and Llama 3.2-3b following our parameter-tuning approach outlined in Section \ref{sec:LM-story}. This evaluation was performed over a subset of 10 software documents left out of the training data (i.e., held-out). With the SFT process (see Section \ref{sec:LM-story}), the Llama3.1-8b model scored 0.065 higher than the base model (0.766 F1), and the Llama3.2-3b model showed a 0.454 increase (0.751). For Llama3.1-8b, we observe the highest gain in the F1 score for action types (0.1), while the F1 score for purposes did not improve from the base model. Our parameter tuned Llama3.1-8b still do not match the performance of Llama3.3-70b, however, these findings show that SFT alone has potential to increase models' performance for the task of privacy behaviors extraction. Further policy alignment of these models (i.e., models with SFT+TPO) reduced accuracy, particularly for the Llama3.2-3b model. These results indicate our dataset does not have enough size and diversity to support these training techniques. A manual review of LLM responses from these models indicates the policy alignment approach led to overfitting \cite{xiao2025foundations}, where models tend to repeat patterns found within the training data. 

All data, code, and experiments are available on our GitHub repository. \footnote{\url{https://anonymous.4open.science/r/privacy_stories_generator/}}

\section{Discussion, Limitations and Research Plan}
\label{sec:discussion}

\subsection{Research Questions' Discussion}
Overall, we found that LLMs are able to extract privacy behaviors and generate privacy stories from software documents. 
Our framework enables LLMs to generate privacy stories by emulating human annotators at inference time. We show that their accuracy can be enhanced through parameter-tuning.

For \textbf{RQ1}, we found that top base LLMs are able to extract privacy behaviors from software documents with up to 0.766 F1 score. Our manual analysis of generated privacy stories revealed that Llama-3.3-70b was able to identify the majority of privacy stories in our ground-truth dataset. Parameter-tuning showed significant improvements, particularly for Llama3.2-3b (with a 0.454 increase in F1) over the test data. All LLMs we tested were able to identify the majority of privacy behaviors found in our dataset, either through initially learning the task at inference time or further parameter-tuning. Our evaluation of LLM responses for extracted privacy behaviors and stories over a diverse set of documents revealed various strengths and weaknesses among models. GPT-4o was more conservative than open-source models, with higher recall scores. For overall F1 score, Llama-3.3-70b excelled at software requirements \& code specification and README documents, GPT-4o excelled in user and developer guides, and Qwen-2.5-32b achieved a 0.051 higher F1 score than any other model for highly structured architecture and DB design diagrams. 
 
For \textbf{RQ2}, our manual evaluation of privacy stories generated by Llama-3.3-70b reveals a weakness that this model tends to hallucinate labels describing privacy behaviors. For example, rather than using labels that were found in our manual annotations, the LLMs may provide labels outside of our taxonomy, often matching the language of the documents. With a larger dataset of software documents, we may use these insights to guide models towards expanding the PAcT taxonomy during annotations. Despite this, failure to follow prompt-based instructions highlights a weakness in providing structured privacy stories based on the standardized labels. 
Our analysis shows that structuring requirements in privacy story templates adds additional challenges for LLMs' accuracy and highlights the need for parameter-tuning approaches to align current models for this task. 

Our manual review of privacy extracted behaviors demonstrates that Llama3.3-70b fails to recognize some behaviors in our taxonomy; instead, it provides more precise labels than we had instructed. We notice all LLMs have the lowest F1 score for data types, with none scoring above 0.641 (see Table \ref{tab:f1_overall-behaviors}).
The performance of models in identifying privacy behaviors shows that GPT4o balances precision and recall. This could indicate that this larger model may be better at following our prompt instructions to use PAcT labels. Lower precision (more false positives) was a trait of all smaller open-source models. This aligns with the findings of our manual evaluation of generated privacy stories; these models often use too many labels in their response to our prompt template, potentially expanding beyond the granularity of the labels found in PAcT. With our SFT approach, both Llama models (3b and 8b) improved more in precision than in recall, with the 8b model achieving a similar balance and overall F1 score to GPT4o.  

\subsection{Limitations}

We acknowledge the \textit{limitations} of our approach. First, gathering software documents that describes applications' interactions with personal data is a challenging task which
limits our ability to perform parameter-tuning to increase LLMs' performance for real-world scenarios. The data we collected is from applications that provide single software documents created in various phases of SDLC, and not only those that are created prior to development. Given this limitation, our result may not transfer to real-world cases where the documents are created in the early phases of SDLC and do not fully resolve the challenge of considering privacy early on in the SDLC.

Second, our approach provides LLMs with snippets or individual files of software documents, where the performance may not transfer to generating privacy requirements from large sets of software documentation. To extend our framework for these cases, we plan to use Retrieval Augmented Generation (RAG) \cite{wang2024astute, xiao2025foundations} in the future. 


Third, our evaluation of generated privacy stories was limited in scope. To avoid annotators' fatigue, only privacy stories generated by Llama3.2-70b were manually reviewed. In future iterations, we will expand this review to multiple models, particularly those that have undergone the parameter-tuning approaches. This evaluation could be further validated by using outside professional reviewers, who can judge the usefulness of LLM responses in scenarios that mimic their development processes. 

\subsection{Research Plan}

There are several main focuses for our future research. First, we will \textit{expand the software documents dataset}. To accomplish this, we will employ NLP tools for classifying and reviewing datasets of software documents. Second, more data can support \textit{further model-tuning approaches}. Through iterative policy tuning, models may improve accuracy \cite{wu2024thinking}, which we plan to perform. We will also test our parameter-tuning approaches on larger models and investigate how their performance can be optimized for generating privacy requirements. Third, we will \textit{further evaluate privacy requirements generated by LLMs} through a user study with developers. 

In this work, we investigate common prompting and post-training techniques for using these LLMs to automate the task of generating privacy requirements given diverse examples of software documents, mimicking human experts. Our preliminary results show promise in using these approaches and provide insights regarding the strengths and weaknesses of various techniques depending on the task. We hypothesize that LLMs, alongside other machine learning tools, can be integrated into various requirement engineering tasks by dynamically generating and evaluating various nonfunctional requirements. Using application code~\cite{jain2023towards,morales2024large}, artifacts, and documentation, such tools could show promise in requirements elicitation throughout the SDLC. In the future, we plan to extend our approach to other types of non-functional requirements, especially those related to security and AI ethics. 

Today, top LLM providers train their models to fit into agentic frameworks where LLM outputs are parsed to call tools, and control precise contexts from various sources, automating more complex tasks \cite{gulli2025agents, anthropic2024effective}. These frameworks can support LLMs to orchestrate various machine learning tools, where, given access to application code, artifacts, and documentation may dynamically support the generation and evaluation of nonfunctional requirements, and more broadly support developers in requirements engineering.



\section{Conclusion}
\label{sec:conclusion}

In this paper, we developed a novel LLM-based framework to generate privacy requirements in the form of privacy stories given software documents as input. We curated a dataset of 25 open-source software documents annotated with 171 privacy behaviors and 93 privacy stories. We extended a taxonomy of privacy behaviors with 3 actions, 50 data types, and 26 purpose labels. Using our prompt templates, we found base LLMs scored as high as 0.766 for the task of extracting privacy behaviors. Following parameter tuning on our dataset, Llama3.2-3b and Llama3.1-8b showed significant (0.454 and 0.065 increase in F1) improvement over a subset of our documents. 

In the future, we will gather and annotate a larger set of software documents to support parameter training techniques, particularly for policy tuning to improve the results. Additionally, we plan to expand our evaluation process with expert opinions over a larger dataset, including models parameter-tuned for the task. We will evaluate LLMs' performance at generating technical and policy-based solutions given an application's privacy stories. We hypothesize that our approach can be extended to first generate privacy stories from documentation, which are then used to deliver personalized solutions for developers to improve the practices of their application in handling private data.

\section*{Acknowledgment}

This research was funded by NSF Award \#2238047.

\bibliographystyle{IEEEtran}
\bibliography{references}


\end{document}